# Optimal Decentralized Economical-sharing Criterion and Scheme for Microgrid

Zhangjie Liu, Mei Su, Yao Sun, *Member, IEEE*, Lang Li and Hua Han

***Abstract*—In order to address the economical dispatch problem in islanded microgrid, this letter proposes an optimal criterion and two decentralized economical-sharing schemes. The criterion is to judge whether global optimal economical-sharing can be realized via a decentralized manner. On the one hand, if the system cost functions meet this criterion, the corresponding decentralized droop method is proposed to achieve the global optimal dispatch. Otherwise, if the system does not meet this criterion, a modified method to achieve suboptimal dispatch is presented. The advantages of these methods are convenient, effective and communication-less.**

## I. INTRODUCTION

MICROGRID has been identified as a key component of modern electrical systems to facilitate the integration of renewable generation units. Economic dispatch, in a way, is considered as one of the core problems in microgrid research. Due to the advantages of high reliability and communication-less, decentralized optimal methods have attracted more and more attention.

To reduce operation cost, some decentralized methods based on droop concept are proposed in [1]-[5]. The nonlinear droop schemes based on equal cost principle are introduced in [1]-[2], and the main idea is to make the cost of each distributed generator (DG) equal by droop control. As a result, the cheaper DGs can produce more power, and the system cost is reduced. Moreover, a nonlinear droop construction approach is proposed based on polynomial approximation [3], which can lower the total cost by selecting the appropriate coefficients. In order to realize plug-and-play, an improved droop construction approach is proposed in [4], which can function independently for each DG by optimizing each one against hypothetical DGs. However, the methods in [1]-[4] are suboptimal. To realize the global optimization by decentralized method, a nonlinear λ-consensus algorithm is introduced in [5]. The incremental costs are subtly embedded into the droop control. When the frequency is synchronized, the total cost is minimized in terms of the equal incremental cost principle (EICP).

Yet, the EICP works only under the convex cost function, which motivates us to consider some more generalized problems: A) under what conditions can the global optimization be realized in decentralized manner? B) How do we design the droop-based controller to achieve the optimization? C) When global optimization cannot be realized by a decentralized manner, how do we design the appropriate suboptimal controller?

The main work of this letter is summarized as:

- An optimal criterion is presented to judge whether global optimal economical-sharing can be realized via a decentralized manner.
- An optimal decentralized economical-sharing scheme is proposed when the criterion is met.
- A suboptimal scheme is proposed when the criterion is not met.

## II. PROBLEM FORMULATION

Mathematically speaking, the objective of economic dispatch problem is to minimize the total generation cost subject to the demand supply constraints as well as the generator constraints. Its model is formulated as

$$\begin{cases} \min \sum_{i=1}^{n} C_i(P_i) \\ s.t.\ \sum_{i=1}^{n} P_i = P_L;\ 0 \le P_i \le P_{i,\max},\ i \in \{1,2,\cdots,n\} \end{cases} \tag{1}$$

where $C_i(P_i)$ is the general comprehensive cost of the $i^{\text{th}}$ DG including maintenance cost, fuel cost, environment cost, renewable energy subsidies, and so on. So, the cost function may be not quadratic or convex.

In AC microgrid, the economic dispatch is usually achieved via the methods based on droop control [1]-[5]. The droop-based control scheme ($P$-$f$) is presented as

$$f_i = f^* - F_i(P_i) \tag{2}$$

where $f_i$, $f^*$ and $P_i$ denote the actual frequency, reference frequency and output active power of the $i^{\text{th}}$ DG, respectively.

Then, when the system achieve frequency synchronization, the following results is obtained

$$F_1(P_1) = F_2(P_2) = \cdots = F_n(P_n),\ \left(0 \le P_i \le P_{i,\max}\right) \tag{3}$$

where $P_{i,\max}$ is the maximal power of the $i^{\text{th}}$ DG. From the balance of power supply-demand, it yields

$$P_1 + P_2 + \cdots + P_n = P_L \quad \left(0 \le P_L \le P_{L,\max}\right) \tag{4}$$

where $P_L$, $P_{i,\max}$ are the total load and the supremum of load, respectively. Then, in steady state, the out power of each DG yields

$$\begin{cases} F_1(P_1) = F_2(P_2) = \cdots = F_n(P_n) \\ \sum_{i=1}^{n} P_i = P_L;\ \ 0 \le P_i \le P_{i,\max},\ i \in \{1,2,\cdots,n\} \end{cases} \tag{5}$$

Normally, the function $F_i(P_i)$ is monotonic to guarantee the uniqueness of the operation point.

Because the constraint in (1) is compact and the cost functions are all continuous, it follows from the well-known extreme value theorem that the economic dispatch problem has a global optimal solution $P^* \triangleq \left(P_1^*, P_2^*, \cdots, P_n^*\right)$.

Then, for the problem of droop-based economic dispatch, the key is how do we design the droop functions $F_i(P_i)$ such that $P^*$ is the solution of problem (5), $\forall P_L \in [0, P_{L,\max}]$.

## III. Proposed Decentralized Economical-Sharing Criterion and Scheme

### *A. The Decentralized Global Optimization Criterion*

For any $P_L \in [0, P_{L,\max}]$, there is a global optimal solution $\left(P_1^*, P_2^*, \cdots, P_n^*\right)$ such that the cost is minimal. Then the optimal solution $P_i^*$ can be regarded as the function of the total load $P_L$. For convenience, we define it as optimal solution function (OSF). Note that $g_i(P_L) = P_i^*(P_L)$ and its implicit condition is expressed as

$$\sum_{i=1}^{n} g_i\left(P_L\right) = P_L \tag{6}$$

The main results of this letter is expressed as follow:

**Theorem 1**. If OSFs are all strictly monotonically increasing, the global optimization can be achieved by taking $F_i(P_i) = m g_i^{-1}(P_L)$, where $m$ is the gain coefficient to keep the frequency deviation in the acceptable range.

**Proof**. When the system reaches the steady state, according to (3) and (4), it yields

$$\begin{cases} g_1^{-1}\left(P_1\right) = g_2^{-1}\left(P_2\right) = \cdots = g_n^{-1}\left(P_n\right) \\ \sum_{i=1}^{n} g_i\left(P_L\right) = P_L; P_{i,\min} \le P_i \le P_{i,\max}, i \in \{1, 2, \cdots, n\} \end{cases} \tag{7}$$

Assume $g_1^{-1}(P_1) = g_2^{-1}(P_2) = \cdots = g_n^{-1}(P_n) = k$, then, we have $P_i = g_i(k)$ and $\sum_{i=1}^{n} g_i(k) = P_L$. Since function $\sum_{i=1}^{n} g_i$ is strictly monotonically increasing, according to (6), we have $k = P_L$. Thus, the solution of (7) is the global optimization $P^*$.

**Theorem 2.** If OSFs are not all strictly monotonically increasing, then the global optimization can not be achieved in decentralized manner.

**Proof.** Assume that there exist $P_{L,1}, P_{L,2} \in (0, P_{L,\max})$ such that $g_k(P_{L,1}) = g_k(P_{L,2})$. We define $(x_{11}, x_{21}, \dots, x_{n1})$ and $(x_{12}, x_{22}, \dots, x_{n2})$ as the solution of (5) when the load is $P_{L,1}$ and $P_{L,2}$, respectively. Thus, it yields

$$\begin{cases} F_1\left(x_{11}\right) = \cdots = F_k\left(x_{k1}\right) = \cdots = F_n\left(x_{n1}\right) \\ F_1\left(x_{12}\right) = \cdots = F_k\left(x_{k2}\right) = \cdots = F_n\left(x_{n2}\right) \\ x_{11} + x_{21} + \cdots + x_{k1} + \cdots + x_{n1} = P_{L,1} \\ x_{12} + x_{22} + \cdots + x_{k2} + \cdots + x_{n2} = P_{L,2} \end{cases} \tag{8}$$

Since $g_k(P_{L,1}) = g_k(P_{L,2})$, namely, $x_{k1} = x_{k2}$, (9) is obtained

$$F_1\left(x_{11}\right) = \cdots = F_n\left(x_{n1}\right) = F_k\left(x_{k1}\right) = F_k\left(x_{k2}\right) = F_1\left(x_{12}\right) = \cdots = F_n\left(x_{n2}\right) \tag{9}$$

Because $P_{L,1} \neq P_{L,2}$, according to (9), there must exist $x_{m1} \neq x_{m2}$ such that $F_m(x_{m1}) = F_m(x_{m2})$, which is contradicted with the precondition that $F_i$ is monotonic. The proof is accomplished.

Then, question A and B is answered by **Theorem 1** and **2.** And the **Optimization Criterion** is summarized as

The global optimization can be achieved in decentralized manner if and only if OSFs are all strictly monotonically increasing for $P_L \in [0, P_{L,\max}]$.

### *B. The Decentralized Suboptimal Scheme*

If OSFs are not all strictly monotonically increasing, the global optimization can not be achieved by decentralized method. However, how do we construct a satisfied suboptimal method? According to the criterion, Fig. 1 implies modified smooth curves instead of the initial non-monotonic curves in order to make the whole curve continuous and monotonous. Define $\gamma_i(P_i)$ as the suboptimal solution function (SOSF). To ensure satisfactory suboptimal solutions, the dotted line is designed according to the following principles

$$\begin{cases} \min \int_{\beta_i}^{\alpha_i} \left(F_i - \gamma_i\right)^2 dP_i \\ s.t. \dfrac{d\gamma_i}{dP_i} \ge \varepsilon; \sum_{i=1}^{n} \gamma_i = P_L \end{cases} \tag{9}$$

where $[\alpha_i, \beta_i]$ is the interval to be fitted and $\varepsilon$ is an appropriate positive scalar. Thus, the economical droop law is constructed as

$$f_i = f^* - m\gamma_i^{-1}\left(P_i\right) \tag{10}$$

Although $\gamma_i(P_i)$ is suboptimal, it has good performances. When $P_L$ belongs to the subintervals such that $g_i(P_i) = \gamma_i(P_i)$ for all DGs, the global optimization can still be achieved. When $P_L$ belongs to the other subintervals, processing (9) ensures that the operation point is close to the optimal point. Thus, the suboptimal droop-based method is also effective and convenient. Question C is answered.

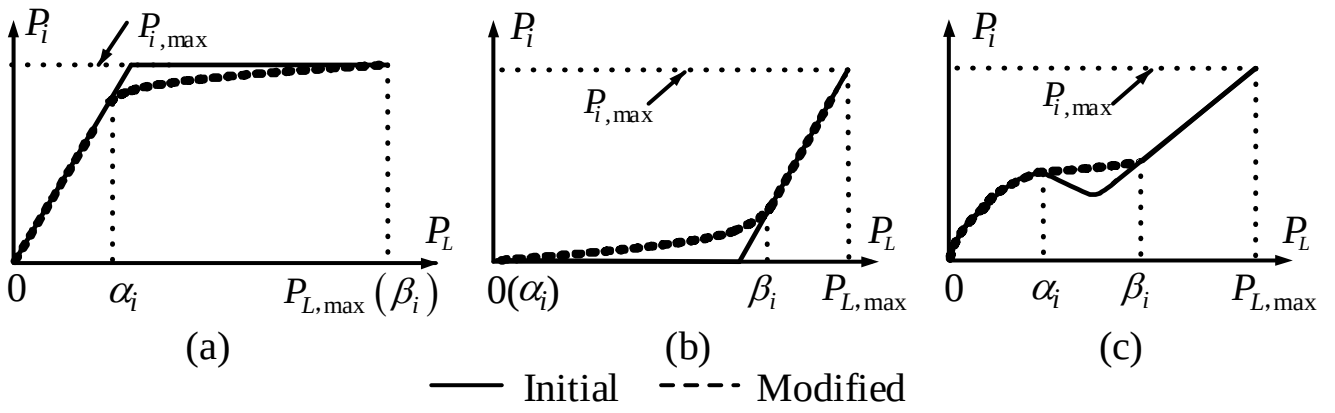

Fig. 1. Construction of suboptimal solution for illustration. (The solid and dotted lines represent OSFs and SOSFs, respectively.)

## IV. Simulation and Experiment

To verify the correctness and effectiveness of the proposed method, two simulation cases (case 1, case 2) and one experiment case (case 3) are tested. The general operation cost function is $C_i\left(P_i\right) = a_i P_i^3 + b_i P_i^2 + c_i P_i + d_i \exp\left(e_i P_i\right)$, and the coefficients and system parameters are listed in Table I.

The OSFs and SOSFs of all cases are obtained by optimization toolbox of MATLAB, which are shown in Fig.3 (a). The $P$-$f$ droop curves in Fig. 3 (b) are obtained according to (10). The number sequences in Fig.3 (a) and (d) are theoretical optimal and operation points, respectively. According to the optimization criterion, if the OSFs are all strictly monotonically increasing, the operation point of the proposed method is optimal. Otherwise, it is suboptimal. The simulation and experimental results are shown in Fig. 3(c), (d) and Fig. 2.

**Case 1**: Fig 3 (a-1) shows that all the OSFs are strictly monotonically increasing, which implies that the global optimization can be achieved. Moreover, Fig 3 (a-1) and (d-1) show that all the operation points coincide with the theoretical global optimal point. Therefore, the cost is minimized and the correctness of **Theorem 1** is verified.

**Case 2 and 3**: Fig. 2 (a-2) implies that the global optimization can be achieved when load value belong to (16, 25]. Otherwise, the operation point is suboptimal. Comparing the operation points with the theoretical global points, we can find that they are completely coincident when load is 20kW and the former is close to the latter when load is 10kW and 15kW. The similar conclusions can also be derived from case 3.

Thus, the correctness and effectiveness of the proposed methods are verified.

Table I
Cost Coefficient for Simulation and Experimental

| DG | a/b/c/d/e $10^{-3}/10^{-3}/10^{-2}/10^{-3}/10^{-1}$ | $P_{max}$ kW | $P_L$ kW |
|---|---|---|---|
| DG1 of case1 | 0/4/0.4/3/2.86 | 10 | 10→15→20 |
| DG1 of case2 | 0.4/-5/6/0/0 | 10 | 10→15→20 |
| DG2 of case1,2 | 0/5.4/0.4/2/2.86 | 10 | 10→15→20 |
| DG3 of case1,2 | 0/3.3/1.1/1/2.86 | 8 | 10→15→20 |
| DG4 of case1,2 | 0/2.4/0.8/4/2.86 | 8 | 10→15→20 |
| DG1 of case3 | 0/800/4/2/28.6 | 1 | 0.8→1.2→1.5 |
| DG2 of case3 | 0/240/8/2/28.6 | 1 | 0.8→1.2→1.5 |

## V. Conclusions

In this letter, the problem whether the global optimal economical dispatch can be realized by *P-f* droop control is studied. The criterion and the corresponding optimal droop control methods are proposed. The salient feature of this criterion is that the OSFs are obtained by off-line calculation. Moreover, the optimal and suboptimal schemes are simple and convenient. The effectiveness of the proposed method has been verified through the simulations and experiments.

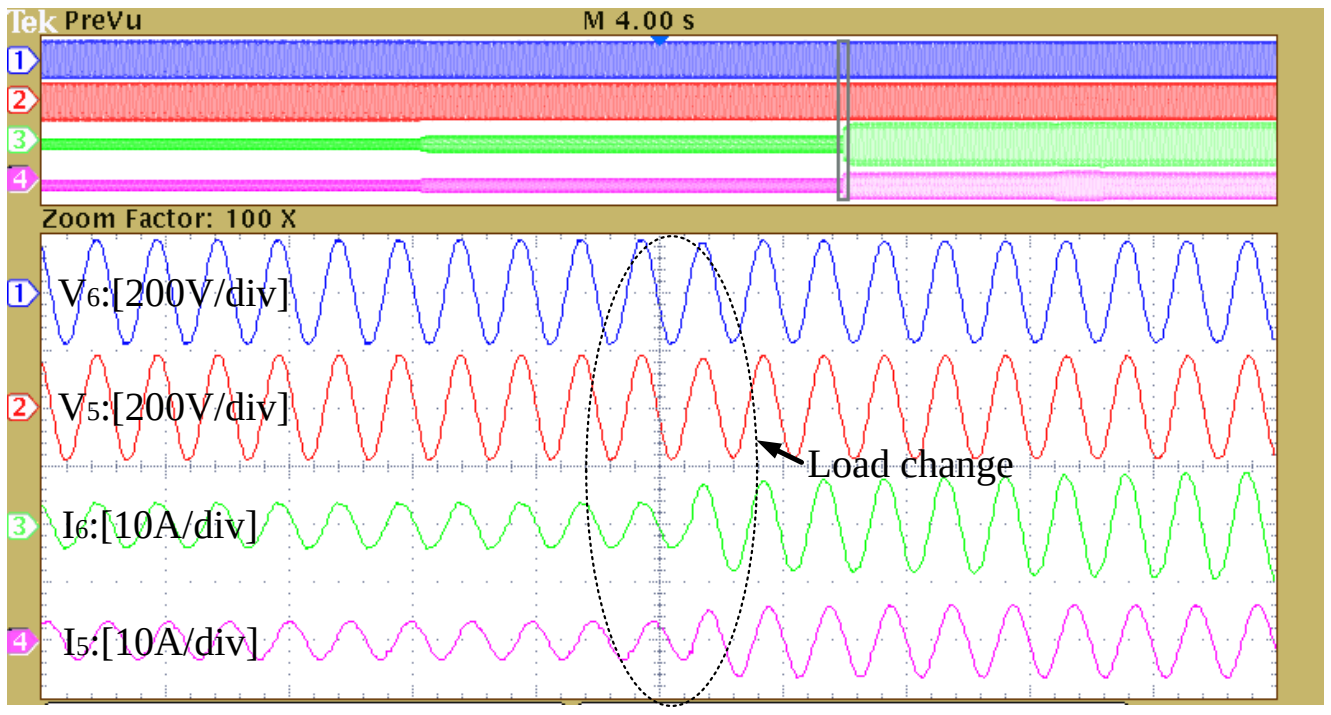

Fig.2. Experimental results of the proposed scheme.

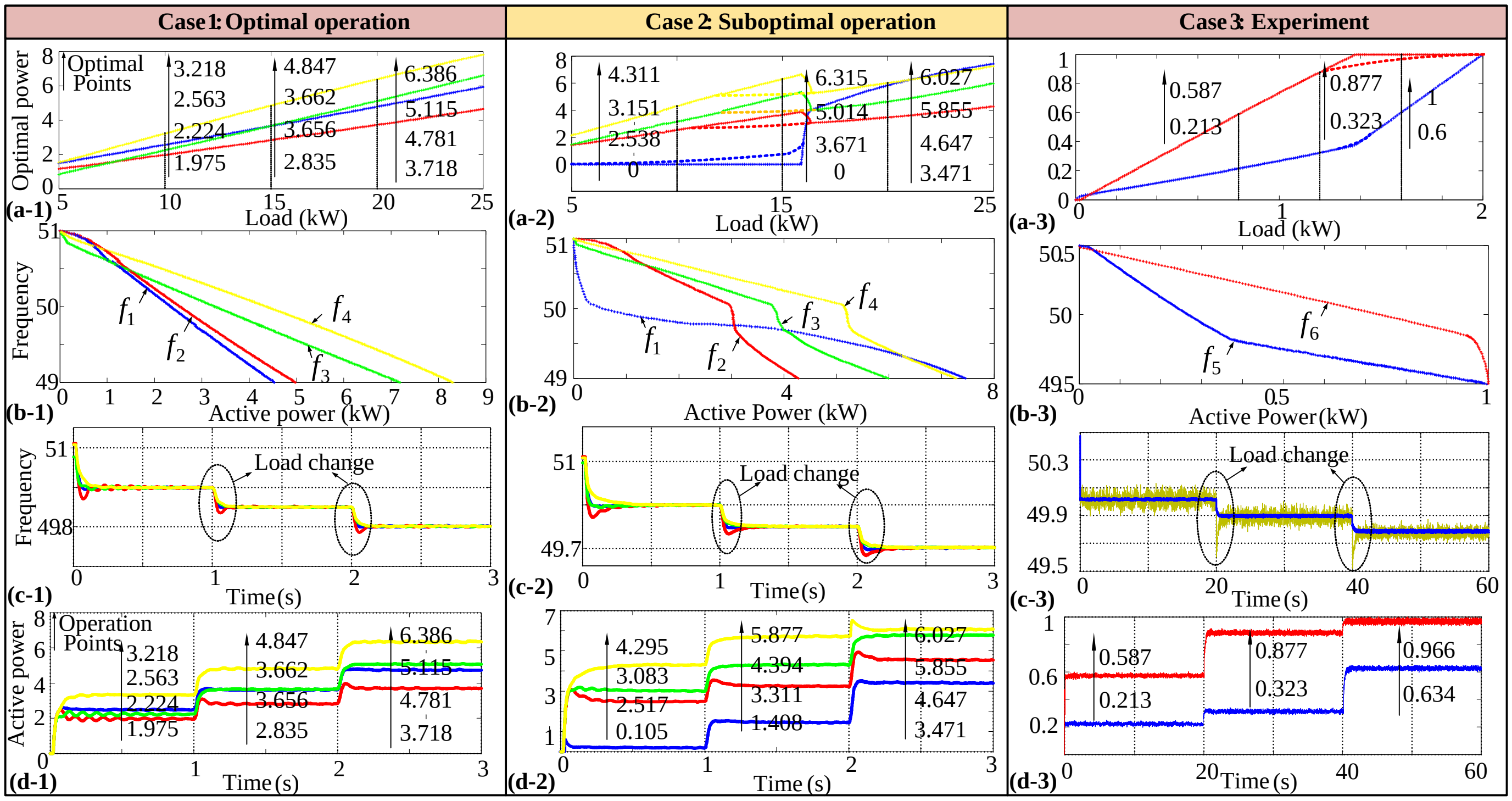

Fig. 3. Simulation and experiment results of (a) OSF $g_i(P_i)$ and SOSF $\gamma_i(P_i)$, (b) droop curve, (c) frequency, (d) active power under three cases.